\documentclass[12pt]{iopart}

\usepackage{iopams}                 
\usepackage{graphicx}
\usepackage{dcolumn}
\usepackage{bm}
\usepackage[dvipsnames]{xcolor}     
\usepackage{mathrsfs}
\usepackage{setstack}               

\newcommand{\ket}[1]{\left|#1\right\rangle}
\newcommand{\bra}[1]{\left\langle #1\right|}
\newcommand{\braket}[1]{\left\langle #1 \right\rangle}

\begin{document}
\date{\today}
\title[Spherical model with non-reciprocal interactions]{\bf Zero-temperature dynamics of the spherical model with non-reciprocal interactions}

\author{Daniel A. Stariolo} 
\address{Universidade Federal Fluminense, Instituto de Física, Av. Litor\^anea s/n, Campus
  da Praia Vermelha, 24210-346 Niter\'oi, RJ, Brazil}
\author{Fernando L. Metz}
\address{Universidade Federal do Rio Grande do Sul, Instituto de Física, 91501-970 Porto Alegre, RS, Brazil}

\begin{abstract}     
 We analytically solve the zero-temperature dynamics of the spherical model with non-reciprocal random interactions drawn
from the real elliptic ensemble of random matrices, where a single parameter $\eta$ continuously interpolates between purely
symmetric ($\eta=1$) and purely antisymmetric ($\eta=-1$) couplings.
 We show that the two-time correlation and response functions depend on both times in the presence of non-reciprocal interactions, reflecting the
 breakdown of time-translation invariance and the absence of equilibrium at long times.
 Nevertheless, the long-time relaxation of the two-time
 observables is governed by exponential decays, in contrast to the slow, power-law relaxation characteristic of the model with purely symmetric interactions.
 We further show that, when the interactions present antisymmetric correlations of strength $\eta <0$, there is a time scale $\tau(\eta)$ above which the dynamics
 undergoes a transition to an oscillatory regime where
 the two-time observables display periodic oscillations with an exponentially decaying amplitude.
 Overall, our results give a detailed account of the dynamics of the spherical model with non-reciprocal interactions at zero temperature, providing
 a benchmark for the study of complex systems with nonlinear and asymmetric interactions.
\end{abstract}

\maketitle
\tableofcontents

\section{Introduction}

In many complex systems, including neural networks \cite{Crisanti2018,Marti2018}, ecosystems \cite{Bunin2017,Roy2019,Altieri2021}, and
epidemic spreading \cite{Satorras2015}, the dynamics of $N$ interacting degrees of freedom $S_1,\ldots S_N$ is described by a 
set of ordinary differential equations of the form:
\begin{equation}
    \frac{dS_i}{dt} = \sum_{j=1}^N J_{ij}\,g_{ij}(S_i,S_j),  \hspace{1cm}  i=1, \ldots N,
    \label{eq:odes}
\end{equation}
where the matrix elements $\{ J_{ij} \}_{i,j=1}^{N}$ define a random network of
pairwise coupling strengths, and the model-dependent functions $\{ g_{ij}(S_i,S_j) \}_{i,j=1}^N$ specify the nature of the interactions between units.
In general, $g_{ij}(S_i,S_j)$ is nonlinear and then solving the
dynamics defined by Eq. (\ref{eq:odes}) for finite $N$ is a formidable task. Significant progress has been
achieved in the limit $N \to \infty$ using dynamical mean-field theory~\cite{Crisanti2018,Roy2019,Altieri2021,Pham2024,Metz2025}.

A simplified yet nontrivial model is obtained when $g_{ij}(S_i,S_j) = S_j$ ($i \neq j$)
and the degrees of freedom satisfy the spherical constraint $\sum_{i=1}^N S_i^2 = N$. This spherical
model~\cite{BookDominicis} has been extensively studied in the context of spin glasses and, more generally, glassy systems.
Beyond its relevance for studying glassy dynamics~\cite{Cugliandolo2002}, the spherical model
provides a simplified framework to address questions related to the nonequilibrium dynamics of complex systems, including the
influence of the interaction matrix  
on entropy production~\cite{Fyodorov2025} and the decay of correlations around a trivial fixed-point~\cite{Marti2018}.
Apart from the global spherical constraint, the time evolution of the model is governed by linear differential equations
that can be solved exactly using results from random matrix theory~\cite{LivanBook}.
The analytic solution for symmetric Gaussian interactions in the $N \to \infty$ limit was first obtained
in the seminal work of~\cite{Cugliandolo1995}. Although the equilibrium properties of the spherical model
with symmetric couplings do not display replica symmetry breaking, which underlies the complexity of spin glasses, the
dynamical evolution of the two-time correlation and response functions still exhibit the phenomenon of aging~\cite{Cugliandolo1995,Henkel2025},
a distinctive feature of glassy dynamics. The standard phenomenology of aging includes the breakdown of time-translation invariance
and the slow, scale-invariant relaxation of two-time observables, typically described by a power-law decay.
In the aging regime, the relaxation timescale grows with the waiting time elapsed since the preparation of the sample and, in the
$N \to \infty$ limit, the system never reaches equilibrium. In models with symmetric couplings, where detailed balance holds, aging effects
are attributed to the complex topology of the energy landscape~\cite{Kurchan1996}, which contains many saddle-points with small curvature that
can trap the dynamics for long times.

Recently, there has been renewed interest in the properties of many-body systems with non-reciprocal, 
asymmetric, or non-Hermitian interactions, both in classical~\cite{Ivlev2015,Fruchart2021,Dinelli2023,Martorell2023,Blumenthal2024,Ros2023,Bhattacherjee2024,Garnier-Brun2024,Hanai2024,Lorenzana2024,Lorenzana2025,Badalotti2025,Mahadevan2025,Guislain2025,Lorenzana2025A}
and quantum settings~\cite{Ashida2020,Zeng2022,Okuma2023,Kennedy2023,BeggHanai2024,Cai2025}. Non-reciprocity is particularly 
relevant for modeling the interactions in complex systems, leading to dynamical behaviours that have no equilibrium
counterpart, including the emergence of oscillations and even chaotic dynamics~\cite{Sompolinsky1988,Fruchart2021,Blumenthal2024,Metz2025}.
The spherical model offers an ideal framework to explore how the relaxation properties of glassy systems
are affected by the introduction of non-reciprocal interactions. In a seminal work, Crisanti and
Sompolinsky~\cite{Crisanti1987} studied the dynamics of the spherical model with a tunable correlation
between $J_{ij}$ and $J_{ji}$, allowing for a systematic analysis of the role of asymmetric coupling strengths.
The main result was that any degree of non-reciprocity destroys the spin glass phase at finite temperatures. Moreover, at
finite but small temperatures, the dynamics exhibits a characteristic
timescale, dependent on the degree of asymmetry in the interactions, below which the system behaves as a frozen
spin glass and above which it behaves as a paramagnet~\cite{Crisanti1987}. A similar behaviour has been also observed in other spin-glass models, where the 
crossover from an aging regime to a time-translation invariant dynamics at long timescales has been termed interrupted aging
~\cite{Cugliandolo1997,Iori1997,Marinari1998,Berthier2012}.

Although long time aging effects are absent at finite temperatures, it was found in~\cite{Crisanti1987} that the model undergoes a transition to
a frozen spin-glass state at zero temperature, where the correlation function remains equal to one.
However, the results of reference~\cite{Crisanti1987} rely on the assumption of time-translation invariance
of the two-time correlation and response functions, which by construction excludes any aging behaviour and, more generally, prevents
the study of nonequilibrium dynamics.
Consequently, the zero-temperature dynamics
of the model remains poorly understood, and it is unclear whether the two-time observables exhibit aging.

Building on known results for the eigenvector correlations of non-Hermitian
random matrices~\cite{Chalker1998,Mehlig2000,Marti2018,Wurfel2024,Crumpton2025}, we analytically solve the zero-temperature dynamics of
the spherical model with non-reciprocal interactions. We derive the long-time behaviour of the correlation and response functions
and show that time-translation invariance is broken for any nonzero contribution of the symmetric part of the interactions, as the two-time observables depend
explicitly on both times.
Similarly to the purely symmetric case, the correlation function displays an initial plateau, whose duration increases with both the waiting time and
the strength of the symmetric couplings.
However, although the model never reaches equilibrium, it does not exhibit slow dynamics: the long-time behaviour of
the correlation and response functions is dominated by exponential decays.
Furthermore, we show that the asymptotic dynamics undergoes a transition to an oscillatory regime as a function of the parameter controlling the correlation
between $J_{ij}$ and $J_{ji}$. For sufficiently strong antisymmetric interactions, the two-time observables exhibit
periodic oscillations with an exponentially decaying amplitude. We compute exactly the period and phase of these oscillations
as functions of the waiting time and the correlation parameter.
These results close a long-standing gap in our understanding of how non-reciprocal interactions impact the zero-temperature dynamics of the spherical model.

The paper is organized as follows. In section~\ref{model} we revisit the definition of the spherical model with non-reciprocal interactions.
In section~\ref{solution}, we present the analytic solution of the dynamics
for arbitrary system sizes $N$. In section~\ref{results}, we discuss our main results for the dynamics of the correlation and response functions
in the $N \to \infty$ limit. Section~\ref{conclusion} summarizes our main findings and outlines possible directions for future work.
In the Appendix we show the derivation of the expressions for the correlation and response for finite-size systems.


\section{The spherical model with non-reciprocal interactions}
\label{model}

The model is defined in terms of $N$ real-valued spins $S_i(t)$ ($i=1,\dots,N$) that evolve in time according to the
zero-temperature Langevin equations~\cite{Cugliandolo1995}:
\begin{equation}
  \frac{\partial S_{i}(t)}{\partial t} = \sum_{j=1}^N S_j(t) J_{ji} - z(t) S_{i}(t) + h_i(t), \hspace{1cm} i=1,\ldots,N,
  \label{hds}
\end{equation}
where $h_i(t)$ are local external fields. The function $z(t)$ ensures that the global spherical constraint:
\begin{equation}
\sum_{i=1}^N S_{i}^{2} (t) = N
\end{equation}  
is satisfied at any time $t$. Thus, the global states of the system lie on the surface of an $N$-$1$-dimensional hypersphere of radius $\sqrt{N}$.

The element $J_{ij} \in \mathbb{R}$ of the $N \times N$ interaction matrix $\boldsymbol{J}$ quantifies the strength of the
directed coupling $i \rightarrow j$, that is, the influence of spin $S_i(t)$ on $S_j(t)$. The diagonal elements of $\boldsymbol{J}$ are set to zero.
To study how the introduction of non-reciprocal
couplings affects the dynamics, we consider interactions
drawn from the real elliptic ensemble of random matrices \cite{Sommers1988}, where the off-diagonal entries of $\boldsymbol{J}$
are identically distributed real Gaussian random variables with mean zero and second moments given by:
\begin{equation}
  \langle J_{ij}^2 \rangle = \frac{1}{N}, \hspace{2cm} \langle J_{ij} J_{ji}  \rangle = \frac{\eta}{N}.
  \label{kopo}
\end{equation}  
The parameter $\eta \in [-1,1]$ controls the degree of correlation between $J_{ij}$ and $J_{ji}$ for any distinct pair $(i,j)$ of sites. For $\eta=0$, $J_{ij}$ and
$J_{ji}$ are statistically independent. For $\eta \rightarrow 1$, the interactions are symmetric, i.e., $J_{ij}=J_{ji}$.
For $\eta \rightarrow -1$, $J_{ij}=-J_{ji}$ and we obtain an anti-correlated ensemble of Gaussian random matrices.

In this work, we analytically compute the two-time correlation 
\begin{equation}
C_N(t,t^{\prime}) = \frac{1}{N} \sum_{i=1}^N S_i(t) S_i(t^{\prime}) 
\label{corrr}
\end{equation}  
and the response function
\begin{equation}
G_N(t,t^{\prime}) = \frac{1}{N} \sum_{i=1}^N \frac{\delta S_{i}(t)}{ \delta h_i(t^{\prime})}\Bigg{|}_{h=0}
\end{equation}  
at time $t$ due to a perturbation at time $t^{\prime} < t$. In particular, we aim to understand the fate of the aging
behaviour when the interactions become non-reciprocal 
($\eta < 1$), since aging is a distinctive feature  of the non-equilibrium
dynamics of the  model with symmetric couplings \cite{Cugliandolo1995}.



\section{Analytical solution of the model dynamics}
\label{solution}

In the following, we derive extact expressions for the two-time autocorrelations and response functions of the spherical model with
asymmetric couplings at zero noise.
The results for arbitrary size $N$ hold for arbitrary interaction matrices and initial conditions, setting a starting point
for analysing the behaviour of the model dynamics in a wide range of situations.
By taking the $N \rightarrow \infty$ limit, we derive analytical expressions for the two-time quantities
when the interactions are sampled from the real elliptic ensemble. We focus on systems prepared on
random initial conditions, with a focus on the fate of the aging behaviour observed in the symmetric case \cite{Cugliandolo1995,Kurchan1996}.

\subsection{Correlation and response for finite-size systems}

The linear Eqs. (\ref{hds}) can be solved by decoupling the dynamical variables in terms
of the eigenvalues and eigenvectors of the interaction matrix $\boldsymbol{J}$. Because the interactions can be
non-reciprocal, $\boldsymbol{J} \neq \boldsymbol{J}^{\dagger}$, the column (right) eigenvectors
of $\boldsymbol{J}$ are not the Hermitian transpose of the row (left) eigenvectors.
Thus, $\boldsymbol{J}$  has a set of right eigenvectors, $\{ \ket{R_\alpha} \}_{\alpha=1,\dots,N}$, and a
set of left eigenvectors, $\{ \bra{L_\alpha} \}_{\alpha=1,\dots,N}$, which fulfill the equations:
\begin{equation}
\boldsymbol{J} \ket{R_\alpha} = \lambda_{\alpha} \ket{R_\alpha}, \hspace{1cm} \bra{L_\alpha} \boldsymbol{J}  = \lambda_{\alpha} \bra{L_\alpha},
\end{equation}
where $\lambda_1,\dots,\lambda_N$ are the complex eigenvalues of $\boldsymbol{J}$.
The left and right eigenvectors form a biorthogonal set 
\begin{equation}
  \braket{L_\alpha|R_\beta} = \delta_{\alpha \beta},
  \label{gu1}
\end{equation}  
and fulfill the completeness relation
\begin{equation}
  \sum_{\alpha=1}^N \ket{R_\alpha}  \bra{L_\alpha} = \boldsymbol{I},
  \label{gu2}
\end{equation}  
where $\boldsymbol{I}$ is the $N \times N$ identity matrix.
The eigenvectors within each set are not orthogonal to each other, i.e., $\braket{L_\alpha|L_\beta} \neq 0$
and $\braket{R_\alpha|R_\beta} \neq 0$. Moreover, Eqs. (\ref{gu1}) and (\ref{gu2}) remain
invariant under the rescalings $\ket{R_\alpha} \rightarrow c_\alpha \ket{R_\alpha}$ and $\bra{L_\alpha} \rightarrow c_{\alpha}^{-1} \bra{L_\alpha}$, with $c_\alpha \in \mathbb{C}$.
This invariance implies that the norm of the left or right eigenvectors can be chosen arbitrarily. Consequently, the dynamical equations for the physical
observables must be invariant under such rescaling.

By introducing the global state vector $\bra{S(t)} = (S_1(t) \,\, S_2(t) \, \dots \, S_N(t))$, Eq. (\ref{hds}) can be written as:
\begin{equation}
  \frac{\partial \bra{S(t) }}{\partial t} = \bra{S(t)} \boldsymbol{J} - z(t) \bra{S(t)}  + \bra{h(t)},
  \label{huasa}
\end{equation}  
with $\bra{h(t)} = (h_1(t) \,\, h_2(t) \, \dots \, h_N(t))$. Let us define the 
projections $r_{\alpha}(t) \in \mathbb{C}$ and $l_{\alpha}(t) \in \mathbb{C}$ of  $\bra{S(t)}$ onto
the eigenvectors:
\begin{equation}
  r_{\alpha}(t) = \braket{S(t) | R_\alpha } \quad \mbox{and} \quad l_{\alpha}(t) = \braket{S(t) | L_\alpha }.
\end{equation}  
By multiplying Eq. (\ref{huasa}) on the right by $\ket{R_\alpha}$ and $\ket{L_\alpha}$, it is straightforward to write
down dynamical equations for  $r_{\alpha}(t)$ and $l_{\alpha}(t)$:
\begin{eqnarray}
  \fl
  \frac{\partial r_\alpha (t) }{\partial t} &=&  \left( \lambda_\alpha - z(t) \right) r_\alpha (t) + h_{\alpha}^{(R)} (t), \hspace{1cm} \alpha=1,\ldots, N, \label{guat1} \\
  \fl
  \frac{\partial l_\alpha (t) }{\partial t} &=&  \sum_{\beta=1}^N  \lambda_{\beta} \braket{L_{\beta} | L_{\alpha} } r_\beta (t)  - z(t) l_\alpha (t) + h_{\alpha}^{(L)} (t), \hspace{1cm} \alpha=1,\ldots, N,
  \label{guat2}
\end{eqnarray}  
where $h_{\alpha}^{(R)} (t)$ and $h_{\alpha}^{(L)} (t)$ are the projections of the external field:
\begin{equation}
h_{\alpha}^{(R)} (t) = \braket{h(t) | R_\alpha } \quad \mbox{and} \quad h_{\alpha}^{(L)} (t) = \braket{h(t) | L_\alpha }.
\end{equation}  
Note that, contrary to the symmetric case, the set of dynamical equations depend not only on the eigenvalue spectrum of
$\boldsymbol{J}$ but also on the set of right and left eigenvectors. 
In terms of $r_{\alpha}(t)$ and $l_{\alpha}(t)$, the correlation function reads:
\begin{equation}
  C_N(t,t^{\prime}) = \frac{1}{N} \sum_{\alpha=1}^N  r_{\alpha}(t) l_{\alpha}^{*}(t^{\prime}),
  \label{udaoe}
\end{equation}
while the response function is given by
\begin{equation}
G_N(t,t^{\prime}) = \frac{1}{N} \lim_{h \rightarrow 0} \sum_{\alpha=1}^N  \sum_{\beta=1}^{N} \frac{\delta r_{\alpha}(t) }{ \delta h_{\beta}(t^{\prime})}
\braket{L_{\alpha} | \beta  },
  \label{kop2}
\end{equation}
where $\{  \ket{\beta} \}_{\beta=1,\dots,N}$ is the standard site basis, with $(\ket{\beta})_{\gamma} = \delta_{\beta \gamma}$. 
As expected, Eqs. (\ref{udaoe}) and (\ref{kop2}) are invariant under
the rescalings $\ket{R_\alpha} \rightarrow c_\alpha \ket{R_\alpha}$
and $\bra{L_\alpha} \rightarrow c_{\alpha}^{-1} \bra{L_\alpha}$.

By solving Eqs. (\ref{guat1}) and (\ref{guat2}) and then setting $h(t)=0$, one can show that the correlation and response functions are given by:
\begin{equation}
  \fl
  C_N(t,t^{\prime}) = \frac{  W_N(t,t^{\prime})   }{\sqrt{ W_N(t,t)  W_N(t^{\prime},t^{\prime})    }}
  \label{toq1}
\end{equation}  
and
\begin{equation}
  \fl
  G_N(t,t^{\prime}) = \sqrt{\frac{W_N(t^{\prime},t^{\prime} ) }{W_N(t,t ) }} \frac{1}{N} {\rm Tr} \left[  e^{\boldsymbol{J} \left( t -  t^{\prime} \right)  } \right]
  -  \frac{1}{N} \frac{ \left[ W_N(t^{\prime},t^{\prime}) \right]^{\frac{1}{2} } }{ \left[ W_N(t,t)\right]^{\frac{3}{2} }  } W_N(t,2 t - t^{\prime}),
       \label{toq2}
\end{equation}  
where the two-time function $W_N(t,t^{\prime})$ reads:
\begin{equation}
  W_N(t,t^{\prime}) = \frac{1}{N^2} \sum_{\alpha \beta = 1}^N
  e^{\lambda_{\alpha} t + \lambda_{\beta}^{*} t^{\prime}} r_{\alpha}(0) r_{\beta}^{*}(0)  \braket{L_{\alpha} | L_{\beta} }.
  \label{juwa}
\end{equation}
The details of the calculations are left to the Appendix. Equations (\ref{toq1}), (\ref{toq2}) and (\ref{juwa}) yield the
relaxation properties of the response and correlation functions for finite $N$, arbitrary
interaction matrix $\boldsymbol{J}$, and arbitrary initial conditions. By numerically diagonalizing $\boldsymbol{J}$,
these equations allow us to study finite size effects on the dynamics of the model. In this work, we will present numerical
results for finite $N$, only to illustrate the approach of the dynamics to its $N \to \infty$ limit, which
is the main focus of the present study.


\subsection{Correlation and response in the thermodynamic limit}

The function $W_N(t,t^{\prime})$ is the central quantity that governs the dynamics of $C_N(t,t^{\prime})$ and $G_N(t,t^{\prime})$.
Let us assume that, in the
limit $N \rightarrow \infty$, $W_N(t,t^{\prime})$ is self-averaging with respect to the randomness in both the initial condition and
the coupling strengths. In mathematical terms, this assumption reads:
\begin{equation}
  \fl
W(t,t^{\prime}) = \lim_{N \rightarrow \infty} N W_N(t,t^{\prime}) = \lim_{N \rightarrow \infty}  \frac{1}{N} \sum_{\alpha \beta = 1}^N
  e^{\lambda_{\alpha} t + \lambda_{\beta}^{*} t^{\prime}} \overline{r_{\alpha}(0) r_{\beta}^{*}(0)}  \braket{L_{\alpha} | L_{\beta} },
\end{equation}  
where $\overline{(\dots)}$ represents an average over the initial conditions $\bra{ S(0) }$. From now on, we consider random
initial conditions in which $\bra{ S(0) }$ is drawn from an
uniform probability distribution subject to the constraint $\braket{ S(0) | S(0) } = N$, with
\begin{equation}
\overline{S_{i}(0)} = 0 \quad \mbox{and} \quad \overline{S_{i}(0) S_{j}(0) } = \delta_{ij}.
\end{equation}  
By explicitly averaging over $\bra{ S(0) }$, we obtain:
\begin{equation}
W(t,t^{\prime}) = \lim_{N \rightarrow \infty} \frac{1}{N} \sum_{\alpha \beta=1}^N
  e^{\lambda_{\alpha} t + \lambda_{\beta}^{*} t^{\prime}  } \braket{L_\alpha |  L_\beta } \braket{R_\beta |  R_\alpha }.
  \label{huaq}
\end{equation}  
If the interaction matrix is symmetric, then $\lambda_{\alpha} = \lambda_{\alpha}^{*}$ and
$\braket{L_\alpha |  L_\beta } = \braket{R_\beta |  R_\alpha } = \delta_{\alpha \beta}$.
Thus, as previously observed, the main difference between symmetric and asymmetric interactions is that in the former case
the behavior of $W(t,t^{\prime})$ is independent
of the eigenvectors, whereas for asymmetric couplings it depends on both the left and right eigenvectors.
This difference arises from the non-orthogonality of the left and right eigenvectors within their respective spaces.

Equation  (\ref{huaq}) depends on the $N \times N$ matrix of eigenvector overlaps,
\begin{equation}
O_{\alpha \beta} = \braket{L_\alpha |  L_\beta } \braket{R_\beta |  R_\alpha }, \hspace{2cm}  \alpha,\beta=1,\dots,N,
\end{equation}  
which was originally introduced in \cite{Chalker1998}.
By using the Dirac-$\delta$ distribution $\delta^2(z) = \frac{1}{2} \delta({\rm Re} z) \delta({\rm Im} z)$ in the complex plane, we can
rewrite Eq.  (\ref{huaq}) as:
\begin{equation}
  W(t,t^{\prime}) = \frac{1}{4} \int d^{2} z_1  \int d^{2} z_2 \ D(z_1,z_2) \  e^{z_1 t + z_{2}^{*} t^{\prime} },
  \label{popo22}
\end{equation}  
where $d^{2}z = d z d z^{*} = 2 \, d {\rm Re} z \, d {\rm Im} z$. The two-point function $D(z_1,z_2)$ is defined as:
\begin{equation}
D(z_1,z_2) = \lim_{N \rightarrow \infty} \frac{4}{N} \sum_{\alpha,\beta =1}^N O_{\alpha \beta} \,
\delta^2(z_1 - \lambda_{\alpha}) \delta^2(z_2 - \lambda_{\beta}).
\end{equation}  
It quantifies the correlations between eigenvectors at two distinct points, $z_1$ and $z_2$, in the complex plane.
In the limit $N \rightarrow \infty$, Eqs. (\ref{toq1}) and (\ref{toq2}) converge to:
\begin{equation}
  C(t,t^{\prime}) = \frac{  W(t,t^{\prime})   }{\sqrt{ W(t,t)  W(t^{\prime},t^{\prime})    }}
  \label{toq111}
\end{equation}  
and
\begin{equation}
G(t,t^{\prime}) = \sqrt{\frac{W(t^{\prime},t^{\prime} ) }{W(t,t ) }} \int_{-\infty}^{\infty} d x \, d y \ \rho(x,y)
\ e^{\left( x + i y \right) \left( t -  t^{\prime} \right)},
       \label{toq222}
\end{equation}  
where $\rho(x,y)$ is the spectral density of $\boldsymbol{J}$ at $z=x + i y$. Equations (\ref{toq111}) and (\ref{toq222})
govern the zero-temperature dynamics of the correlation and response
functions in the thermodynamic limit for arbitrary interaction matrices $\boldsymbol{J}$ and random
initial conditions. The functions $C(t,t^{\prime})$ and $G(t,t^{\prime})$ are entirely
determined by the spectral properties of $\boldsymbol{J}$ through $D(z_1,z_2)$ and $\rho(x,y)$.

Let us now present the equations for $C(t,t^{\prime})$ and $G(t,t^{\prime})$ corresponding to the elliptic random-matrix model for the non-reciprocal interactions.
For random interaction matrices $\boldsymbol{J}$ sampled from the real Gaussian elliptic
ensemble, defined through Eq. (\ref{kopo}), the eigenvalues are uniformly distributed inside an ellipse in the complex plane,
and the spectral density reads \cite{Girko1986,Sommers1988}
\begin{equation}
  \rho(x,y) = \frac{1}{\pi (1 - \eta^2)} \ \Theta \left[1 - \left( \frac{x^2}{(1+\eta)^2} +  \frac{y^2}{(1-\eta)^2}   \right)  \right],
  \label{specel}
\end{equation}  
with $\Theta(\dots)$ denoting the Heaviside step function.
Moreover, the analytic expression for the two-point eigenvector correlator $D(z_1,z_2)$ for the complex elliptic ensemble is given by \cite{Chalker1998,Marti2018}
  \begin{equation}
D(z_1, z_2) = \frac{ (1 + \eta^2)z_1 z_{2}^{*}  - (1-\eta^2)^2 - \eta \left[ z_{1}^2 + (z_{2}^{*})^2 \right] }{\pi^2 (1 - \eta^2) |z_1 - z_2|^{4}} \quad (z_1 \neq z_2),
  \end{equation}  
for $z_1$ and $z_2$ inside the ellipse in the complex plane.
Without loss of generality, from now on we set $t^{\prime} = t_w$ and $t=t_w + \tau$, where $t_w$ is the waiting time and $\tau$ the elapsed time. Building
on the random-matrix analytic results of \cite{Sommers1988,Chalker1998,Marti2018}, we obtain 
\begin{equation}
  C(t_w + \tau,t_w) = \frac{W(t_w + \tau,t_w)}{\sqrt{W(t_w + \tau) W(t_w) }}
  \label{jj0}
\end{equation}  
and
\begin{equation}
  G(t_w + \tau,t_w) = \sqrt{\frac{W(t_w)}{W(t_w + \tau)}} \frac{I_1\left( 2 \sqrt{\eta} \, \tau   \right) }{\sqrt{\eta} \, \tau},
  \label{gg0}
\end{equation}  
where $I_n (x)$ is the modified Bessel function of the first kind. The two-time function $W(t_w + \tau,t_w)$ is given by
\begin{eqnarray}
  \fl
  W(t_w + \tau,t_w) &=& (1 + \eta^2 ) I_0 \left[ \psi \left(t_w , \tau   \right)   \right] - 2 \eta \left[ 1 + \frac{ 2 (1 - \eta)^2 \tau^2 }{ \left(  \psi \left(t_w , \tau   \right)   \right)^2  }  \right]
  I_2 \left[ \psi \left(t_w , \tau   \right)   \right] \nonumber \\
  &-& \frac{1}{t_w (t_w + \tau)} \sum_{k=1}^{\infty} \eta^k k^2 I_{k} \left( 2 t_w \sqrt{\eta} \right)  I_{k} \left( 2 ( t_w + \tau) \sqrt{\eta} \right),
  \label{jj2}
\end{eqnarray}  
with
\begin{equation}
  \psi \left(t_w , \tau   \right) = 2 \sqrt{\eta \tau^2 + (1 + \eta)^2 t_w (t_w + \tau )  }.
  \label{guga56}
\end{equation}  
The one-time function $W(t) = W(t,t)$ follows by setting $\tau=0$ in Eq. (\ref{jj2}):
\begin{equation}
  \fl
  W(t) = (1 + \eta^2) I_0 \left( 2 (1 + \eta) t   \right) - 2 \eta I_2 \left( 2 (1 + \eta) t   \right) - \frac{1}{t^2} \sum_{k=1}^{\infty} \eta^{k} k^{2} I_{k}^{2} \left( 2 t \sqrt{\eta} \right).
  \label{jj111}
\end{equation}  
Equations (\ref{jj0}) and (\ref{gg0}) enable to perform a detailed analysis of how the symmetry
parameter $\eta$ influences the dynamics of the spherical model defined in Eq. (\ref{hds}). 

Setting $\eta=1$ in Eq. (\ref{jj2}) and using properties of the Bessel functions, we obtain
\begin{equation}
 W(t_w + \tau,t_w) = \frac{I_{1}\left[ 2 \left( 2 t_w + \tau   \right) \right]}{2 t_w + \tau } \quad \mbox{and} \quad W(t) =  \frac{I_{1}\left( 4 t \right)}{2 t},
\end{equation}
recovering well-known analytic results for symmetric interactions \cite{Cugliandolo1995}, namely
\begin{equation}
  C(t_w + \tau,t_w) = \frac{ I_{1}\left[ 2 \left( 2 t_w + \tau   \right) \right]   }{\sqrt{I_{1}\left( 4 t_w \right) I_{1}\left[ 4 (t_w + \tau) \right] } }
  \frac{\sqrt{4 t_w \left( t_w + \tau  \right)}}{ (2 t_w + \tau  )  }
  \label{eq:corr-eta-1}
\end{equation}  
and
\begin{equation}
G(t_w + \tau,t_w) = \sqrt{\frac{(t_w + \tau ) \, I_1(4 t_w)  }{t_w \, I_1 \left[ 4 (t_w + \tau  )  \right]   }} \frac{I_1 \left( 2 \tau  \right)  }{ \tau  }.
\end{equation}
These expressions exhibit aging behaviour in the regime $t, t_w \to \infty$, with $0< t_w/t < 1$.
For arbitrary $\eta <1$, the behaviour of the correlation and response is qualitatively different from the symmetric case, as we will show in the next section.


\section{Results}
\label{results}

In this section, we analyze the effect of the symmetry parameter $\eta$ on the dynamics
of $C(t_w + \tau,t_w)$ and $G(t_w + \tau,t_w)$, in the limit $N \rightarrow \infty$. Since the long-time dynamics
is qualitatively different depending on whether $\eta >0$ or $\eta < 0$, we analyze the two regimes separately.

\subsection{Regime $0 \leq  \eta \leq 1$}

\begin{figure}[htbp]
\includegraphics[scale=1.1]{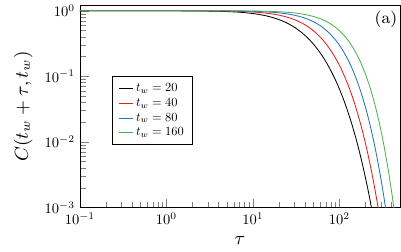}
\includegraphics[scale=1.1]{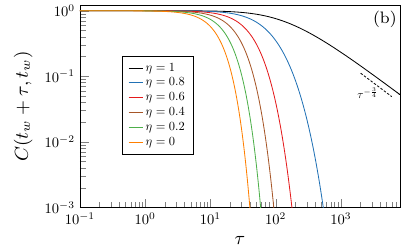}
\caption{Autocorrelation function $C(t_w + \tau,t_w)$ of the spherical model with non-reciprocal interactions drawn from the elliptic  ensemble
      of random matrices  with symmetry parameter $\eta$ (see Eq. (\ref{kopo})). These results are obtained by numerically computing $C(t_w + \tau,t_w)$
      from Eqs. (\ref{jj0}) and (\ref{jj2})-(\ref{jj111}), which are valid in the $N \to \infty$ limit.
(a) $\eta=0.7$ and different values of the waiting time $t_w$. (b) $t_w=40$ and different $\eta$.}
    \label{gugu1}
\end{figure}

Figure \ref{gugu1} shows $C(t_w + \tau,t_w)$ as a function of $\tau$ obtained by numerically
computing the autocorrelation from Eqs. (\ref{jj0}) and (\ref{jj2})-(\ref{jj111}). Since the series in Eq. (\ref{jj2}) converges extremely fast, we numerically compute $W(t_w+\tau,t_w)$ by truncating the series at a sufficiently large $k$. Similarly to the behaviour of the
model with symmetric couplings, $\eta =1$, the behaviour of $C(t_w + \tau,t_w)$ for $0 \leq \eta < 1$ depends on both
times, $t_w$ and $\tau$. For fixed $\eta$ (Figure \ref{gugu1}(a)), the system remains frozen in the initial state and the correlation exhibits
a plateau at $C(t_w + \tau,t_w) = 1$ whose duration increases with $t_w$, until the system finally relaxes
for large enough $\tau$. Fig. \ref{gugu1}(b) illustrates
how $\eta$ affects the relaxation of $C(t_w + \tau,t_w)$ at fixed $t_w$. As $\eta$ decreases, the length of the plateau
also decreases. However, the most pronounced effect of asymmetry in the coupling strengths appears in the decay of $C(t_w + \tau,t_w)$ for large $\tau$. 
While for $\eta=1$ the correlation decays as a power-law, for $\eta < 1$ this decay is much faster, exponential in $\tau$. In summary, the phenomenology
remains qualitatively similar to the symmetric case: the correlation exhibits a plateau for $\tau \ll t_w$, whose duration is controlled by $t_w$ and $\eta$; the system then
leaves the plateau and attains an asymptotic regime for $\tau \gg t_w$, strongly influenced by $\eta$. However, unlike the symmetric case,
for $\eta < 1$ the long-time dynamics is exponential in $\tau$, as will be shown in the following.

Let us now characterize the asymptotic behaviour of $C(t_w + \tau,t_w)$ for large $t_w$ and different regimes of $\tau$.
\begin{itemize}
\item For $\tau/t_w \to 0$, it is straightforward to show from
  Eqs. (\ref{jj0}) and (\ref{jj2})-(\ref{jj111}) that $C(t_w + \tau,t_w) \to 1$. This corresponds to
  the plateau regime shown in Fig. \ref{gugu1}.

\item Consider now the time scales where $t_w \gg 1$ and $\tau \gg 1$, with $0 < t_w/\tau \ll 1$.
This corresponds to the long-time asymptotic regime in Fig. \ref{gugu1}(b), where the decay of the correlation
function strongly depends on the symmetry parameter $\eta$.
For $\eta=1$, one can show from Eq. (\ref{eq:corr-eta-1}) that $C(t_w + \tau,t_w)$ exhibits aging behaviour,
with the power-law decay \cite{Cugliandolo1995}:
  \begin{equation}
  C(t_w + \tau,t_w) \simeq 2 \sqrt{2} \left( \frac{t_w}{\tau} \right)^{\frac{3}{4}} , \hspace{1.0cm} 0<t_w/\tau \ll 1.
  \label{ggbb}
\end{equation}  
For $0 < \eta < 1$, however, the correlation decays in the form:
\begin{equation}
  C(t_{w} + \tau,t_w) \simeq \left[ \frac{(1 + \eta)^2 }{ \eta} \frac{t_{w}}{\tau}  \right]^{\frac{5}{4}} 
  \exp{\left[ -r(\eta) \tau  \right]   },  \hspace{0.7cm} 0<t_w/\tau \ll 1,
  \label{trws}
\end{equation}  
with the symmetry-dependent rate
\begin{equation}
r(\eta) = \left(1 - \frac{(1 + \eta)}{\sqrt{\eta}} \frac{t_w}{\tau}\right)(1 - \sqrt{\eta})^{2}.
\label{eq:c-rate}
\end{equation}  
Equation (\ref{trws}) is derived by expanding Eq. (\ref{guga56}) for 
\begin{equation}
\tau \gg  \frac{(1 + \eta)}{\sqrt{\eta}}\,t_w   ,
\end{equation}
which defines the time scale for the onset of the exponential decay.
Since the limit $\eta \to 0$ of Eq. (\ref{trws})
is singular, the case of $\eta=0$  must be analyzed separately, as will be discussed below.
As $\eta \to 1$, the characteristic time for the decay of correlations,
$1/r(\eta)$, diverges, indicating the breakdown of the exponential regime. 
\end{itemize}

\begin{figure}[htbp]
  \centering
  \includegraphics[scale=1.1]{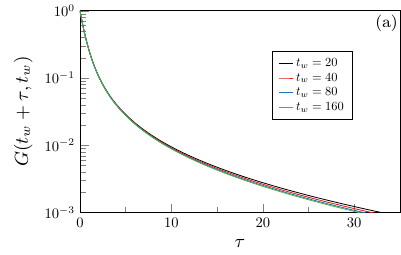}
    \includegraphics[scale=1.1]{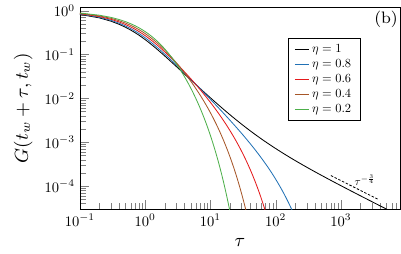}
    \caption{Response function $G(t_w + \tau,t_w)$ of the spherical model with non-reciprocal interactions drawn from the elliptic  ensemble
      of random matrices with symmetry parameter $\eta$ (see Eq. (\ref{kopo})). These results are obtained by numerically computing $G(t_w + \tau,t_w)$
      from Eqs. (\ref{gg0}) and (\ref{jj111}), which are valid in the thermodynamic limit.
      (a) $\eta=0.7$ and different values of $t_w$. (b) $t_w=40$ and different $\eta$.}
    \label{gugu2}
\end{figure}

Next, we study the dynamics of the response function $G(t_w + \tau,t_w)$. Figure \ref{gugu2} 
shows $G(t_w + \tau,t_w)$ as a function of $\tau$ obtained by numerically evaluating $G(t_w + \tau,t_w)$ from Eqs. (\ref{gg0}) and (\ref{jj111}). 
For small $\tau$ the curves for different $t_w$ collapse, and $G(t_w + \tau,t_w)$ depends only on $\tau$. For
large $\tau$, the curves for different $t_w$ split, showing that $G(t_w + \tau,t_w)$ exhibits a weak dependence on the waiting time. Fig. \ref{gugu2}(b) illustrates
the strong effect of $\eta$ on the long-time behaviour of $G(t_w + \tau,t_w)$. For symmetric interactions ($\eta=1$) and sufficiently large $\tau$, $G(t_w + \tau,t_w)$ decays as
a power-law, whereas for $0 < \eta < 1$, the response shows an exponential decay for large $\tau$.

%
\begin{itemize}
\item For $\tau/t_w \rightarrow 0$, $G(t_w + \tau,t_w)$ presents a time-translation invariant
  regime,
\begin{equation}
  G(t_w + \tau,t_w) =  e^{-(1 + \eta)\tau} \frac{I_{1}(2 \sqrt{\eta} \tau )  }{\sqrt{\eta} \tau  },
  \label{gfgf1}
\end{equation}
as illustrated in Fig. \ref{gugu2}(a). Setting $\eta=1$ in Eq. (\ref{gfgf1}), we recover the expression in \cite{Cugliandolo1995}.

\item Consider now the time scales in which $t_w \gg 1$ and $\tau \gg 1$, with $t_w/\tau \ll 1$.
  For $\eta=1$,  $G(t_w + \tau,t_w)$ displays the power-law decay \cite{Cugliandolo1995}
\begin{equation}
  G(t_w + \tau,t_w) \simeq \frac{1}{\sqrt{4 \pi}} (t_w \tau)^{-3/4},  \hspace{0.5cm} 0<t_w/\tau \ll 1,
  \label{yutr}
\end{equation}
whereas for $0 < \eta < 1$ we find
\begin{equation}
  G(t_w + \tau,t_w) \simeq \frac{1}{\sqrt{4 \pi }}  ( \eta\,t_{w} \tau)^{-1/4} \, \frac{e^{- (1 - \sqrt{\eta})^{2} \tau}}{\tau}, \hspace{0.5cm} 0<t_w/\tau \ll 1.
  \label{uhtrw}
\end{equation}  
This behaviour is illustrated in Fig. \ref{gugu2}. The absence of an exponential prefactor in $t_w$ (present in the corresponding result for
the autocorrelation, see. Eqs. (\ref{trws}) and (\ref{eq:c-rate})) reflects the weak dependence of $G(t_w + \tau,t_w)$ on the waiting time. 
For $\eta=1$, Eq. (\ref{uhtrw}) reduces to a power law decay with an exponent
different from that in Eq. (\ref{yutr}).
\end{itemize}

Now we discuss the behaviour of the correlation and response functions for $\eta=0$.
In this case, from Eqs. (\ref{jj2}) and (\ref{jj111}), the correlation and response are given by:
\begin{equation}
C(t_w + \tau, t_w) = \frac{I_0 \left(  2 \sqrt{t_w (t_w + \tau)     } \right)}{\sqrt{I_0 \left( 2 t_w   \right) I_0 \left( 2 (t_w + \tau)   \right)  }  } 
\end{equation}
and
\begin{equation}
G(t_w + \tau, t_w) = \sqrt{\frac{ I_0 \left( 2 t_w   \right)  }{ I_0 \left( 2 (t_w + \tau)   \right)  }  }.
\end{equation}
Let us analyze the asymptotic regimes of $C(t_w + \tau, t_w)$ and $G(t_w + \tau, t_w)$ for large $t_w$.
\begin{itemize}
\item For $\tau/t_w \rightarrow 0$, the dynamics is time-translation invariant, with $C(t_w + \tau, t_w) = 1$ and 
$G(t_w + \tau, t_w) = e^{-\tau}$. 
\item On the other hand, when $t_w$ and $\tau$ are both large, such that $0 < t_w/\tau \ll 1$, the correlation and response exhibit the leading behaviours:
\begin{eqnarray}
   C(t_w + \tau, t_w) &\simeq&  e^{-\left( 1 + \frac{2 t_w}{\tau} - 2 \sqrt{\frac{t_w}{\tau}} \right) \tau  },    \\
   G(t_w + \tau, t_w) &\simeq& \left(  \frac{\tau}{t_w}  \right)^{\frac{1}{4}} e^{-\tau},  \hspace{1.0cm} 0<t_w/\tau \ll 1.
\end{eqnarray}
\end{itemize}

\begin{figure}[htbp]
  \centering
  \includegraphics[scale=1.1]{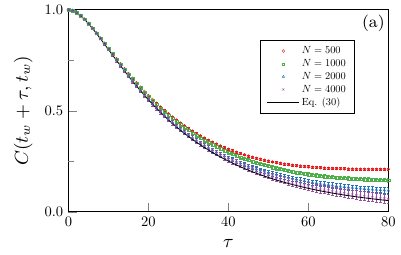}
  \includegraphics[scale=1.1]{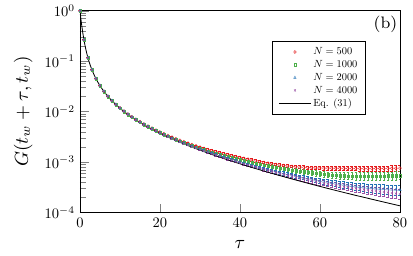}
  \caption{Comparison between numerical results, obtained from Eqs. (\ref{toq1}) and (\ref{toq2}) for different system sizes $N$, and the analytical expressions for the correlation
    and response functions in the limit $N \rightarrow \infty$ for $\eta=0.7$ and $t_w = 10$.
    The error bars in the numerical data represent the standard deviation of the empirical average computed over a large number of independent realizations.}
    \label{gu3}
\end{figure}

We conclude this subsection by comparing our analytic results for $N \rightarrow \infty$, Eqs. (\ref{jj0}) and (\ref{gg0}), with finite-size results derived from Eqs.  
(\ref{toq1}) and (\ref{toq2}).
For finite $N$ one can compute the function $W_N(t,t')$, given by Eq. (\ref{juwa}), by numerically diagonalizing the
interaction matrix $\mathbf{J}$, thus producing the set of eigenvalues and eigenvectors 
$\{\lambda_\alpha, \  \ket{R_\alpha},\  \ket{L_\alpha} \}_{\alpha=1,\dots,N}$. By sampling the initial state from a Gaussian distribution
on the hypersphere, one then computes its projection onto the right
eigenvectors, $r_{\alpha}(0) = \braket{S(0) | R_\alpha }$, which allows to obtain the function $W_N(t,t')$. With these results and after computing the trace 
${\rm Tr} \left[  e^{\boldsymbol{J} \left( t -  t^{\prime} \right)  } \right] = \sum_{\alpha=1}^N e^{\lambda_{\alpha} \left( t -  t^{\prime} \right)  }$, one can determine
the two-time autocorrelation and response functions for fixed $N$ from Eqs. (\ref{toq1}) and (\ref{toq2}), respectively.
Although finite
size effects are significant for large $\tau$, Fig. \ref{gu3} shows that the numerical data systematically approach the
theoretical predictions as $N$ increases, thereby confirming our analytic findings for $C(t_w + \tau, t_w)$ and $G(t_w + \tau, t_w)$ in the $N \to \infty$ limit.

\subsection{Regime $-1 \leq  \eta < 0$}

In this regime the antisymmetric part of the interactions is dominant.
We first consider the limiting case $\eta=-1$ of purely antisymmetric couplings, where Eq. (\ref{guga56}) reduces
to $\psi = 2 i \tau$. Using the property $I_n(ix) = i^n J_n(x)$, with $J_n(x)$ denoting the Bessel function of order $n$, one can show that
Eqs. (\ref{jj0})-(\ref{jj2}) reduce to:
\begin{equation}
  C(t_w + \tau, t_w) = G(t_w + \tau, t_w) = \frac{J_{1}(2 \tau)}{\tau}.
  \label{yy89}
\end{equation}  
For large $\tau$, the leading term of Eq. (\ref{yy89}) reads:
\begin{equation}
  C(t_w + \tau, t_w) = G(t_w + \tau, t_w) \simeq \frac{1}{\sqrt{\pi} \tau^{3/2} } \cos{\left(2 \tau - \frac{3 \pi}{4}    \right) }.
  \label{dada34}
\end{equation}
Thus, for $\eta=-1$, the model exhibits a time translation invariant dynamics with periodic oscillations,
whose amplitude decays as a power-law.

We now focus on the regime $-1 < \eta < 0$. First, we examine the form of the
functions $W(t)$ and $W(t_w + \tau,t_w)$ for negative $\eta$. From Eq. (\ref{jj111}),  we
obtain:
\begin{equation}
  \fl
  W(t) = (1 + \eta^2) I_0 \left[ 2 (1 - |\eta|) t   \right] + 2 |\eta| I_2 \left[ 2 (1 - |\eta|) t   \right] - \frac{1}{t^2} \sum_{k=1}^{\infty} |\eta|^{k} k^{2} J_{k}^{2} \left( 2 t \sqrt{|\eta|} \right).
  \label{jj2232}
\end{equation}  
The two-time function $W(t_w + \tau,t_w)$ depends on $\psi(t_w,\tau)$, Eq. (\ref{guga56}), which can be real or purely imaginary depending on
the relation between $\tau$ and $t_w$. There is a characteristic time:
\begin{equation}
  \tau_{*} = \frac{(1 - |\eta|)}{|\eta|} t_w,
  \label{jkl}
\end{equation}  
obtained from the condition $\psi(t_w,\tau)=0$, which marks the onset of oscillatory dynamics in the correlation function.
For $\tau < \tau_{*}$, the argument of the
square root in Eq. (\ref{guga56}) is positive, and the behaviour of Eq. (\ref{jj2}) is mainly governed by the modified Bessel
functions $I_n(x)$, particularly for large $t_w$. Consequently, $C(t_w + \tau,t_w)$ exhibits a transient, non-oscillatory dynamics for $\tau < \tau_{*}$.
In contrast, for $\tau > \tau_{*}$, the argument in Eq. (\ref{guga56}) becomes negative, $\psi(t_w,\tau)$ is purely imaginary,
and Eq. (\ref{jj2}) takes the form:
\begin{eqnarray}
  \fl
  W(t_w + \tau,t_w) &=& (1 + \eta^2 ) J_0 \left( 2 \sqrt{\phi}  \right) - 2 |\eta| \left[ 1 - \frac{ (1 + |\eta|)^2 \tau^2 }{ 2 \phi }  \right]
  J_2 \left( 2 \sqrt{\phi}  \right) \nonumber \\
  \fl
  &-& \frac{1}{t_w (t_w + \tau)} \sum_{k=1}^{\infty} |\eta|^k k^2 J_{k} \left( 2 t_w \sqrt{|\eta|} \right)  J_{k} \left( 2 ( t_w + \tau) \sqrt{|\eta|} \right),
  \label{jj222}
\end{eqnarray}  
with
\begin{equation}
\phi = | (1 - |\eta|)^2 t_w (t_w + \tau ) -|\eta| \tau^2 |.
\end{equation}  
Equation (\ref{jj222}), given only in terms of Bessel functions $J_{n}(x)$, leads to an oscillatory dynamics of
$C(t_w + \tau,t_w)$ for $\tau > \tau_{*}$. Note that the
characteristic time $\tau_{*}$ shows the limiting behaviours $\tau_{*} \rightarrow 0$ and $\tau_{*} \rightarrow \infty$ for $|\eta| \rightarrow 1$
and $|\eta| \rightarrow 0$, respectively.

\begin{figure}[htbp]
  \centering
  \includegraphics[scale=1.1]{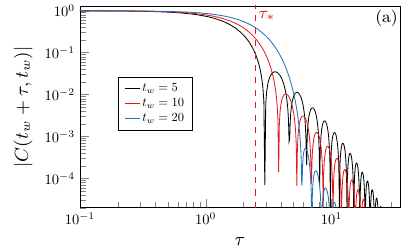}
  \includegraphics[scale=1.1]{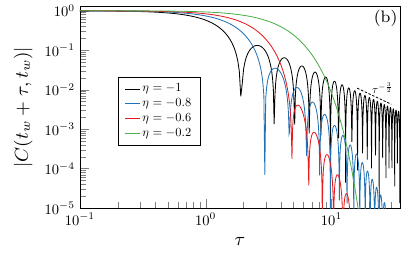}
  \caption{Absolute value of the autocorrelation function of the spherical model with non-reciprocal interactions drawn from the
   real elliptic ensemble of random matrices with symmetry parameter $\eta<0$. These results are obtained by numerically computing
    $|C(t_w + \tau,t_w)|$ from Eqs. (\ref{jj0}), (\ref{jj2232}) and (\ref{jj222}), which are valid in the $N \to \infty$ limit.
    (a) $\eta=-0.8$ and different values of $t_w$. The vertical dashed line indicates the characteristic time $\tau_{*}$, Eq. (\ref{jkl}), which
    marks the onset of oscillations for $t_w=10$. (b) $t_w=5$ and different values of $\eta < 0$.  }
    \label{gugru1}
\end{figure}
Figure \ref{gugru1} shows the relaxation of the absolute value  of the autocorrelation as a function
of $\tau$, obtained from Eq. (\ref{jj0}) for negative $\eta$. We note that, for $-1 < \eta < 0$, the dynamics of $C(t_w + \tau,t_w)$
is not time-translation invariant, as was the case for positive $\eta$ values. The duration
of the short initial plateau, $C(t_w + \tau,t_w) \simeq 1$, slightly increases with $t_w$, while the functional form of $|C(t_w + \tau,t_w)|$ describing
the non-oscillatory decay up to $\tau=\tau_{*}$ seems qualitatively similar to the results for $0 < \eta < 1$ (see Fig. \ref{gugu1}).
For $\tau > \tau_{*}$, the system enters the oscillatory regime, with an amplitude that decays faster than a power-law.
Fig. \ref{gugru1}(b) illustrates how $\eta < 0$ influences the relaxation of $|C(t_w + \tau,t_w)|$. As $|\eta|$ increases, the duration of the plateau
decreases until it disappears at $\eta=-1$, where the characteristic time $\tau_*$ vanishes and the
oscillatory dynamics becomes time-translation invariant. For $\eta=-1$, the amplitude decays as
a power-law, analogous to the slow dynamics observed at the opposite limit $\eta = 1$.
\begin{figure}[h!]
  \centering
  \includegraphics[scale=1.1]{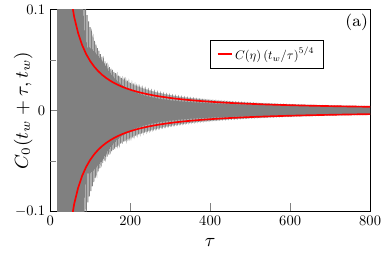}
   \includegraphics[scale=1.1]{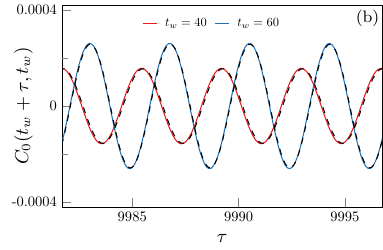}
   \caption{Rescaled correlation function $C_0(t_w + \tau,t_w)$ for negative $\eta$. These results are obtained by numerically computing
 $C_0(t_w + \tau,t_w)$ from Eqs. (\ref{jj0}) and (\ref{bobo7}).
    (a) $C_0(t_w + \tau,t_w)$ for $\eta=-0.7$ and $t_w=40$. The solid red lines show the
analytic result for the amplitude extracted from Eqs. (\ref{uywd}) and (\ref{bobo7}).
(b) Long-time periodic oscillations of $C_0(t_w + \tau,t_w)$ for $\eta=-0.7$
     and two values of $t_w$. The dashed lines are the analytic prediction of Eq. (\ref{uywd}).}
    \label{rara5}
\end{figure}

It is clear from Fig. \ref{gugru1} that the correlation function exhibits a plateau $C(t_w + \tau,t_w) = 1$ for $\tau/t_w \rightarrow 0$.
However, the distinctive feature of the dynamics for negative $\eta$ is the long-time oscillatory behaviour, which we now characterize analytically.
For large $t_w$ and $\tau \gg \tau_{*}$, the correlation decays as:
\begin{equation}
  \fl
C(t_w + \tau,t_w) \simeq C(\eta) \left( \frac{t_w}{\tau} \right)^{\frac{5}{4}} e^{-2(1 - |\eta|)t_w - (1 - |\eta|) \tau}
\cos{\left[ 2 \sqrt{|\eta|} \tau -  \xi(t_w,\eta) \right]},
  \label{uywd}
\end{equation}  
with the phase:
\begin{equation}
  \xi(t_w,\eta) = \frac{(1+\eta)^2}{\sqrt{|\eta|}} t_w + \frac{5 \pi}{4}
  \label{iu90}
\end{equation}  
and the prefactor:
\begin{equation}
C(\eta) = 2 \sqrt{\frac{1+\eta  }{ \sqrt{|\eta|} }} \frac{(1+\eta)^2}{|\eta|}.
\end{equation}
Therefore, for fixed $t_w \gg 1$, the correlation oscillates with period $T = \pi / \sqrt{|\eta|}$, phase $\xi(t_w,\eta)$, and an amplitude
that decreases as $\tau^{-5/4} e^{-(1 - |\eta|) \tau}$. To numerically verify the analytical prediction of Eq. (\ref{uywd}),
it is convenient to introduce the rescaled function:
\begin{equation}
  C_0(t_w + \tau,t_w) = e^{2(1 - |\eta|)t_w + (1 - |\eta|) \tau} C(t_w + \tau,t_w),
  \label{bobo7}
\end{equation}  
where $C(t_w + \tau,t_w)$ is computed from Eq. (\ref{jj0}). Figure \ref{rara5} illustrates the
oscillatory behaviour of $C_0(t_w + \tau,t_w)$, showing that the amplitude follows the power-law
$C(\eta) \left( \frac{t_w}{\tau} \right)^{\frac{5}{4}}$ for sufficiently large $\tau$.  
Fig. \ref{rara5}(b) displays in more detail the long-time periodic dynamics of $C_0(t_w + \tau,t_w)$,
fully confirming Eq. (\ref{uywd}) and the analytic expressions for both the period and the
phase of oscillations.
We note that, as the degree of antisymmetry in the couplings decreases ($|\eta| \rightarrow 0$), the oscillations become slower.
The phase $\xi(t_w,\eta)$ is also determined by the waiting time $t_w$, which sets the initial time from which the correlation is computed.

\begin{figure}[htbp]
  \centering
  \includegraphics[scale=1.1]{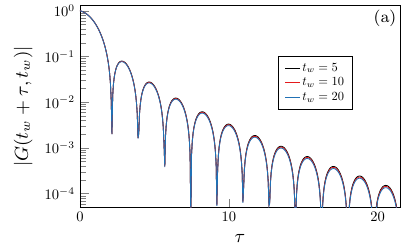}
  \includegraphics[scale=1.1]{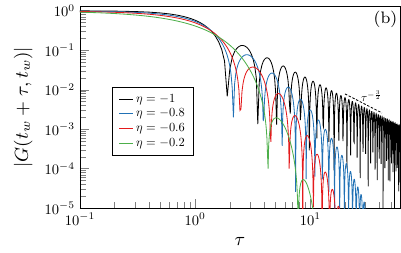}
  \caption{
 Absolute value of the response function of the spherical model with non-reciprocal interactions drawn from the
  real elliptic ensemble with symmetry parameter $\eta<0$. These results are obtained by numerically computing
    $|G(t_w + \tau,t_w)|$ from Eqs. (\ref{gg0}), (\ref{jj2232}) and (\ref{jj222}), which are valid in the $N \to \infty$ limit.
    (a) $\eta=-0.8$ and different $t_w$. (b) $t_w=20$ and different values of $\eta < 0$.}
    \label{gugru2}
\end{figure}

Let us now examine the response function $G(t_w + \tau,t_w)$ for negative $\eta$. Its behaviour is illustrated in figure \ref{gugru2}.
For $\eta=-1$, the dynamics is time-translation invariant and the amplitude of oscillations decays
as a power-law (see Eq. (\ref{dada34})).
Figure \ref{gugru2}(a) shows that, for $-1 < \eta < 0$, $G(t_w + \tau,t_w)$ displays a very weak dependence on $t_w$, 
analogous to its behaviour for $0 < \eta < 1$.
For $\tau/t_w \rightarrow 0$, the response decays as
\begin{equation}
G(t_w + \tau,t_w) = e^{-(1 - |\eta|) \tau} \frac{J_1 \left( 2 \sqrt{|\eta|} \tau  \right)  }{\sqrt{|\eta|} \tau },
\end{equation}  
which corresponds to a time-translation invariant regime. In contrast, for $t_w \gg 1$ and $t_w/\tau \ll 1$, the asymptotic
behaviour of $G(t_w + \tau,t_w)$ depends explicitly on $t_w$,
\begin{equation}
  G(t_w + \tau,t_w) \simeq A(\eta) \, t_{w}^{- \frac{1}{4}} \tau^{- \frac{5}{4} }   e^{-(1 - |\eta|) \tau} \cos{\left(2 \sqrt{|\eta|} \tau - \frac{3 \pi}{4}    \right)},
  \label{popu9}
\end{equation}  
with
\begin{equation}
A(\eta) = \frac{1}{\sqrt{\pi |\eta| \sqrt{|\eta|}  }}.
\end{equation}
Hence, the response exhibits periodic oscillations with a constant phase, independent of $t_w$, at variance with Eq. (\ref{iu90}). The
function $G(t_w + \tau,t_w)$ oscillates with the same period $T = \pi/\sqrt{|\eta|}$ as $C(t_w + \tau,t_w)$, and its
amplitude decays as $\tau^{- \frac{5}{4} }   e^{-(1 - |\eta|) \tau}$ for fixed $t_w \gg 1$. Equation (\ref{popu9}) contains no exponential
prefactor in $t_w$, which explains the weak dependence on $t_w$ observed in Fig. \ref{gugru2}(a).

\begin{figure}[htbp]
  \centering
  \includegraphics[scale=1.1]{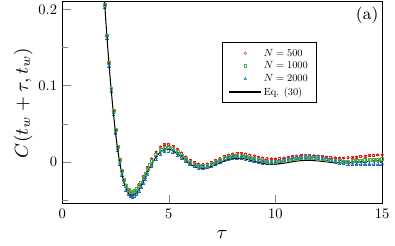}
  \includegraphics[scale=1.1]{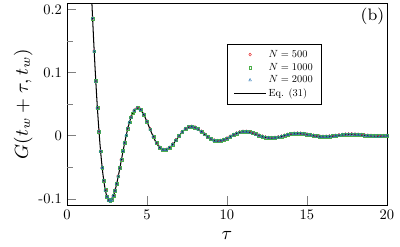}
  \caption{Comparison between  numerical results, obtained from Eqs. (\ref{toq1}) and (\ref{toq2}) for different system sizes $N$, and the analytical expressions for the correlation
    and response functions in the limit $N \rightarrow \infty$, for $\eta=-0.9$ and $t_w = 10$.
    The error bars represent the standard deviation of the empirical average computed over a large number of independent realizations.}
    \label{guop}
\end{figure}

We end this subsection by comparing our analytical expressions for $C(t_w + \tau,t_w)$ and $G(t_w + \tau,t_w)$
with finite-size numerical results in the case of negative $\eta$.
Figure \ref{guop} shows the dynamics of the correlation and response for $N \rightarrow \infty$, together with numerical results
obtained from Eqs. (\ref{toq1}) and (\ref{toq2}). For $\eta < 0$, finite-size effects are very weak, and the numerical data exhibit
an excellent agreement with the analytical predictions of Eqs. (\ref{jj0}) and (\ref{gg0}) for sufficiently large $N$.
%


\section{Summary and discussion}
\label{conclusion}

The spherical model with non-reciprocal interactions
 characterized by a symmetry parameter $-1 \leq \eta \leq 1$  was first studied long ago
by Crisanti and Sompolinsky \cite{Crisanti1987}. Their results showed that introducing asymmetric couplings ($|\eta| < 1$)
suppresses aging at finite temperatures, leading to a time-translation invariant dynamics.
However, the possibility that nontrivial dynamical effects could persist at zero temperature has remained open since then.
In this work, we have addressed this issue and solved the nonequilibrium dynamics of the spherical model with non-reciprocal interactions
at zero temperature, showing that
for random initial conditions the long-time behaviour never settles into a time translation-invariant regime for any $|\eta| < 1$.  

By analytically computing the two-time autocorrelation and
response functions in the thermodynamic limit, we have found that the two-time observables
depend explicitly on both times.
The relaxation of the correlation function is qualitatively similar to the standard
picture observed for purely symmetric interactions \cite{Cugliandolo1995}: for large waiting times $t_w$, defining the elapsed
time since the preparation of the sample, the correlation exhibits an initial plateau whose duration increases with both $t_w$ and $\eta$.
However, unlike the power-law relaxation of the two-time observables at $\eta=1$, the long-time behaviour of the correlation
and response functions of the spherical model is governed, for any $|\eta| < 1$, by an exponential decay with an algebraic prefactor.
This is an important difference of the spherical model with non-reciprocal interactions with respect to the slow dynamics characteristic of the
symmetric case.

We have also shown that the long-time behaviour
of both the correlation and response functions undergoes a transition to an oscillatory regime at $\eta=0$. For $0 \leq \eta < 1$,
the dynamics of the two-time observables does not exhibit any oscillatory behaviour.
For $-1 < \eta < 0$, that is, when the antisymmetric part of the interactions becomes sufficiently strong, the two-time observables
display periodic oscillations beyond a characteristic elapsed time $\tau^*(\eta,t_w)$, with a period and phase determined by $t_w$ and $\eta$.
Recent works \cite{Lorenzana2024,Lorenzana2025,Lorenzana2025A} have shown
that a spherical model composed of two groups of spins globally coupled to each other through antisymmetric interactions
also undergoes a transition to an oscillatory regime. The main difference
between our results and these works lies essentially in the amplitude of oscillations.
While in our case the amplitude of the correlation exhibits an exponential 
decay, in reference \cite{Lorenzana2024} it follows a power-law decay,
characteristic of aging.

The slow dynamics of the spherical model with symmetric interactions is commonly attributed to a
complex energy landscape \cite{Kurchan1996}, in which the number of saddle-points scales with the system size $N$.
It is then natural to expect that the number of stationary states of the model with non-reciprocal interactions
play an important role in its dynamical behaviour.
From Eq. (\ref{eq:odes}), the stationary states, $\bra{S(t)}=\bra{S_{\rm st}}$, satisfy:
\begin{equation}
  \bra{S_{\rm st}} \boldsymbol{J} = z_{\rm st} \bra{S_{\rm st}},
  \label{nu1}
\end{equation}  
where the Lagrange multiplier at the stationary states reads:
\begin{equation}
  z_{\rm st} = \frac{1}{N} \bra{S_{\rm st}} \boldsymbol{J} \ket{S_{\rm st}  }.
  \label{nu2}
\end{equation}  
By normalizing the left eigenvectors of the interaction matrix $\boldsymbol{J}$ as $\braket{L_{\alpha} | L_{\alpha} }=N$, one finds
that Eqs. (\ref{nu1}) and (\ref{nu2}) admit the solutions :
\begin{equation}
\bra{S_{{\rm st},\alpha}} = \bra{L_{\alpha}},  \hspace{1.5cm} \alpha=1,\dots,N,
\end{equation}  
with $z_{{\rm st},\alpha} = \lambda_{\alpha}$. Thus, there are in total $N$ solutions of Eqs. (\ref{nu1}) and (\ref{nu2}), although not all of them
correspond to physical states. Indeed, since the components of $\bra{S_{\rm st}}$ must be real, only those solutions associated with real
eigenvalues $\lambda_{\alpha} \in \mathbb{R}$ represent physical states.
It is known that the average number of real eigenvalues of matrices $\boldsymbol{J}$ drawn from the elliptic ensemble
scales as $\sqrt{N}$ for large $N$ \cite{Forrester2008,Byun2021}.
Thus, as soon as the antisymmetric part of the interactions is introduced, there is a drastic reduction in the number of stationary
states of the model (see also \cite{Fyodorov2016}). These stationary points are expected to be sparsely distributed
over the surface of the hypersphere, which may lead to a faster relaxation compared with the symmetric case.
Studying the stability of these fixed points and their relation to the dynamics of the spherical model is an interesting direction of future work.
This kind of analysis has been done in a related model with higher order interactions~\cite{Fournier2025}. However,
in this last model, the number of equilibria is exponential in $N$, even in the presence of
non-reciprocal interactions, at variance with the behaviour of the model discussed in this work.

Finally, we point out that, in the crossover regime of weak non-Hermiticity \cite{Fyodorov1997}, where the
symmetry parameter scales with $N$ as $\eta = 1 - \mathcal{O}(1/N)$, the number of real eigenvalues of $\boldsymbol{J}$
remains proportional to $N$ ($N \gg 1$) \cite{Crumpton2025}.
This implies that the complexity of the configuration space characterizing the symmetric spherical model remains essentially
unchanged under the introduction of a sufficiently weak antisymmetric perturbation.
Hence, in this weakly non-Hermitian regime, the two-time observables of the spherical model with non-reciprocal interactions
are expected to exhibit slow dynamics.
Understanding the dynamics of the spherical model in this regime is therefore an interesting problem that we will address in
a future work. It would be also important to clarify whether complex systems governed by nonlinear and asymmetric
interactions \cite{Bunin2017,Crisanti2018,Roy2019,Metz2025}, such as neural networks and ecosystems, can display nontrivial
dynamical properties, including the breakdown of time-translation invariance and aging.


\addcontentsline{toc}{section}{Acknowledgements} 
\section*{Acknowledgements} 
F. L. M. acknowledges a visiting researcher fellowship from FAPERJ (Grant No 204.646/2024) and financial
support from CNPq (Grants No 402487/2023-0 and No 310255/2025-2). F. L. M. warmly thanks the hospitality of the Physics Institutes of Universidade do Estado do Rio de Janeiro and Universidade Federal Fluminense, where this work has been developed. D. A. S. acknowledges
financial support from CNPq (Grants CNPq/MCTI 10/2023, 402487/2023-0 and CNPq 4/2021, 307337/2021-9). 

\addcontentsline{toc}{section}{Appendix  \hspace{0.5cm}The correlation and response functions for a finite system}
\section*{Appendix. The correlation and response functions for a finite system}


In this appendix we show the main steps in the derivation of the finite-size expressions for the correlation and response
functions, Eqs. (\ref{toq1}) and (\ref{toq2}), which are valid for arbitrary interaction matrix $\boldsymbol{J}$ and initial conditions.
This is achieved by solving the coupled system of Eqs. (\ref{guat1}) and (\ref{guat2}) for the projections $r_{\alpha}(t) = \braket{S(t) | R_\alpha }$ and $l_{\alpha}(t) = \braket{S(t) | L_\alpha }$.
The solution of Eq. (\ref{guat1}) is given by

\begin{equation}
r_\alpha (t)  = r_\alpha (0) \Gamma_{h}(t)  e^{\lambda_{\alpha} t} +
e^{\lambda_{\alpha} t }
  \Gamma_{h}(t) \int_{0}^{t} d t^{\prime} \frac{h_{\alpha}^{(R)}(t^{\prime})}{ \Gamma_{h}(t^{\prime}) } e^{- \lambda_{\alpha} t^{\prime} },
  \label{yuqa}
\end{equation}  
where we introduced the function
\begin{equation}
\Gamma_{h}(t) = \exp{\left[- \int_{0}^{t} dt^{\prime} z(t^{\prime}) \right] }.
\end{equation}  
Our notation highlights that $\Gamma_{h}(t)$ depends on the external fields $\{ h_i(t) \}_{i=1}^{N}$, which will be important in the calculation of the
response function. Inserting the above solution for $r_\alpha (t)$ in Eq. (\ref{guat2}), we obtain
\begin{equation}
  \frac{\partial l_\alpha (t) }{\partial t} =    - z(t) l_\alpha (t) + F_\alpha(t),
  \label{utae}
\end{equation}  
with
\begin{equation}
  \fl
  F_\alpha(t) =  \sum_{\beta=1}^N  \lambda_{\beta} \braket{L_{\beta} | L_{\alpha} }  \Gamma_h(t)  e^{\lambda_{\beta} t}
 \left[ r_\beta (0) +  \int_{0}^{t} d s \, \frac{h_{\beta}^{(R)}(s)}{ \Gamma_h(s)  }
   e^{- \lambda_{\beta} s  }    \right] 
 + h_{\alpha}^{(L)}(t). \label{udow}
\end{equation}  
Equation (\ref{utae}) is a linear dynamical equation with an additive noise $F_\alpha(t)$ generated by the
time evolution of the amplitudes $r_\alpha (t)$. The overlap $\braket{L_{\beta} | L_{\alpha} }$ in Eq. (\ref{udow}) couples the dynamics of $l_\alpha (t)$ with the
whole set of right eigenvectors, which is a direct consequence of the non-orthogonality of the eigenvectors.
The solution of Eq. (\ref{utae}) reads
\begin{equation}
  l_\alpha (t) = l_{\alpha}(0) \Gamma_h(t) + \Gamma_h(t) \int_{0}^{t} d t^{\prime}  \frac{ F_\alpha(t^{\prime})  }{\Gamma_h(t^{\prime})}.
\end{equation}  
Substituting Eq. (\ref{udow}) in the above expression and using the property $l_{\alpha}(0) = \sum_{\beta=1}^N r_{\beta}(0) \braket{L_{\beta} | L_{\alpha}  }$, we find
\begin{eqnarray}
  \fl
 l_\alpha (t) &=& \Gamma_h(t) \sum_{\beta=1}^N r_{\beta}(0)   \braket{L_{\beta} | L_{\alpha} } e^{\lambda_{\beta} t} + \Gamma_h(t) \sum_{\beta=1}^N  \lambda_{\beta} \braket{L_{\beta} | L_{\alpha} }
 \int_{0}^{t} d t^{\prime}   e^{\lambda_{\beta} t^{\prime}}  U_{\beta}(t^{\prime}|h) \nonumber \\
 \fl
 &+&   \Gamma_h(t) \int_{0}^{t} d t^{\prime}  \frac{h_{\alpha}^{(L)}(t^{\prime})}{ \Gamma_h(t^{\prime}) }, \label{guras}
\end{eqnarray}  
where
\begin{equation}
U_{\beta}(t|h) = \int_{0}^{t} d s \, \frac{h_{\beta}^{(R)}(s)   }{\Gamma_h(s)}  e^{- \lambda_{\beta} s  }  .
\end{equation}  
The solutions for $r_{\alpha}(t)$ and $l_{\alpha}(t)$, Eqs. (\ref{yuqa}) and (\ref{guras}), are determined by the external fields and $\Gamma_h(t)$. Thus, in order
to close the system of equations, we need an equation for $\Gamma_h(t)$. Substituting Eqs. (\ref{yuqa}) and (\ref{guras}) in the spherical constraint
\begin{equation}
\sum_{\alpha=1}^{N} r_{\alpha}(t) l_{\alpha}^{*}(t) = N,
\end{equation}  
we obtain an explicit equation for $\Gamma_h(t)$,
\begin{eqnarray}
  \fl
  \frac{N}{\Gamma_{h}^{2}(t)} &=& \sum_{\alpha \beta=1}^N r_{\alpha}(0)  r_{\beta}^{*}(0)  \braket{L_{\beta} | L_{\alpha} }^{*} e^{\lambda_{\alpha} t + \lambda_{\beta}^* t}
  + \sum_{\alpha \beta=1}^N r_{\alpha}(0) \lambda_{\beta}^{*} e^{\lambda_{\alpha} t} \braket{L_{\beta} | L_{\alpha} }^{*} \int_{0}^t d s e^{\lambda_{\beta}^* s} U_{\beta}^{*} (s|h) \nonumber \\
  \fl
  &+& \sum_{\alpha =1}^N r_{\alpha}(0) e^{\lambda_{\alpha} t}  \int_{0}^t d s \, \frac{\left( h_{\alpha}^{(L)} (s) \right)^{*}  }{\Gamma_{h}(s)}
  +  \sum_{\alpha \beta=1}^N  r_{\beta}^{*}(0)  \braket{L_{\beta} | L_{\alpha} }^{*} e^{\lambda_{\alpha} t + \lambda_{\beta}^* t} U_{\alpha}(t|h) \nonumber \\
  \fl
  &+& \sum_{\alpha \beta =1}^N \lambda_{\beta}^*  e^{\lambda_{\alpha} t} \braket{L_{\beta} | L_{\alpha} }^{*} U_{\alpha}(t|h) \int_{0}^t d s e^{\lambda_{\beta}^* s} U_{\beta}^{*} (s|h) \nonumber \\
  \fl
  &+& \sum_{\alpha =1}^N  e^{\lambda_{\alpha} t} U_{\alpha}(t|h) \int_{0}^t d s \, \frac{\left( h_{\alpha}^{(L)} (s) \right)^{*}  }{\Gamma_{h}(s)}.
  \label{pupu}
\end{eqnarray}  
The solutions of Eqs. (\ref{yuqa}), (\ref{guras}) and (\ref{pupu}) determine the time evolution of the projections, $\{ r_{\alpha}(t) \}_{\alpha=1}^N$ and $\{ l_{\alpha}(t) \}_{\alpha=1}^N$, from which
one can compute the correlation and response functions.

Although we are not interested in the effects of the external fields on the dynamics, the generic form of Eq. (\ref{pupu}) for arbitrary
$\{ h_i(t)  \}_{i=1}^N$ is crucial to compute the response function, given by Eq. (\ref{kop2}).
The derivative $\frac{\delta r_{\alpha}(t) }{ \delta h_i(t^{\prime})} \Big{|}_{h=0}$ of Eq. (\ref{yuqa}) reads
\begin{equation}
  \fl
  \frac{\delta r_{\alpha}(t)}{\delta h_{i}(t^{\prime}) }\Big{|}_{h=0} = r_{\alpha}(0) e^{\lambda_{\alpha} t} \frac{\delta \Gamma_{h} (t) }{\delta h_{i}(t^{\prime}) }\Big{|}_{h=0}
  + \frac{\Gamma_{0} (t) }{\Gamma_{0} (t^{\prime}) } e^{\lambda_{\alpha} (t - t^{\prime})  } \braket{i | R_{\alpha} } \quad (t^{\prime} < t),
  \label{uia}
\end{equation}  
where $\Gamma_{0}(t) = \lim_{h \rightarrow 0} \Gamma_{h}(t)$  fulfills
\begin{eqnarray}
  \frac{N}{\Gamma_{0}^{2}(t)} &=& \sum_{\alpha \beta=1}^N r_{\alpha}(0)  r_{\beta}^{*}(0)  \braket{L_{\beta} | L_{\alpha} }^{*} e^{\lambda_{\alpha} t + \lambda_{\beta}^* t}.
  \label{pupu34}
\end{eqnarray}  
An explicit expression for $\frac{\delta \Gamma_{h} (t) }{\delta h_{i}(t^{\prime}) }\Big{|}_{h=0}$
is obtained from Eq. (\ref{pupu}), namely
\begin{equation}
  \frac{\delta \Gamma_{h} (t) }{\delta h_{i}(t^{\prime}) }\Big{|}_{h=0} = - \frac{1}{N} \frac{\Gamma_{0}^3(t) }{\Gamma_{0}(t^{\prime})}
       {\rm Re} \left[  \sum_{\alpha \beta=1}^N r_{\alpha}(0)  \braket{L_{\alpha} | L_{\beta} } \braket{R_{\beta} | i } e^{\lambda_{\alpha} t + \lambda_{\beta}^{*} (t - t^{\prime})   }    \right].
       \label{bd}
\end{equation}
Combining Eqs. (\ref{kop2}), (\ref{uia}) and (\ref{bd}), we arrive at the following expression 
\begin{eqnarray}
  G_N(t,t^{\prime}) &=& \frac{\Gamma_{0} (t) }{\Gamma_{0} (t^{\prime}) } \frac{1}{N} \sum_{\alpha=1}^{N} e^{\lambda_{\alpha} \left( t -  t^{\prime} \right)  } \nonumber \\
  &-&  \frac{\Gamma_{0}^3(t) }{\Gamma_{0}(t^{\prime})} {\rm Re}
  \left[ \frac{1}{N^2} \sum_{\alpha \beta=1}^N r_{\alpha}(0)  r_{\beta}^{*}(0)  \braket{L_{\alpha} | L_{\beta} } e^{\lambda_{\alpha} t + \lambda_{\beta}^{*} (2 t - t^{\prime})   }    \right]
       \label{bbfs}
\end{eqnarray}  
for the response function of finite-size systems. It is easier to derive an equation for the correlation function $C_N(t,t^{\prime})$ in the absence
of external fields. Setting $h_i(t)=0$ in Eqs. (\ref{yuqa}) and (\ref{guras}), and
substituting them in Eq. (\ref{udaoe}), we get
\begin{equation} 
  C_N(t,t^{\prime}) = \Gamma_{0}(t)\Gamma_{0}(t^{\prime}) 
  \frac{1}{N} \sum_{\alpha \beta =1}^N r_\alpha (0) r_{\beta}^{*} (0) \braket{L_{\alpha} | L_{\beta} } e^{\lambda_{\alpha} t + \lambda_{\beta}^{*} t^{\prime}  }.   \nonumber  
  \label{huwaodp1}
 \end{equation} 
By introducing the two-time function $W_N(t,t^{\prime})$, defined in Eq. (\ref{juwa}), the above expressions for $G_N(t,t^{\prime})$ and $C_N(t,t^{\prime})$ lead
to Eqs. (\ref{toq1}) and (\ref{toq2}) in the main text.


\end{document}